\documentclass[aps,prb,groupedaddress]{revtex4-2}

\usepackage {amssymb}
\usepackage {amsmath}
\usepackage{graphicx}

\begin{document}


\title{Influence of chirality on the electron transmission through step-like potential
in zigzag, armchair, and (2m,m) carbon nanotubes}


\author{Lyuba Malysheva}
\email[]{malysh@bitp.kiev.ua}
\affiliation{Bogolyubov Institute for Theoretical Physics, 03680 Kiev, Ukraine}


\date{\today}

\begin{abstract}

We report the one-electron spectrum and eigenstates of infinite achiral and chiral $(2m,m)$ carbon nanotubes
found by using  the analytic solution to the Schr\"odinger equation for the tight-binding H\"uckel-type Hamiltonian.
With the help of matching the wave functions on the interfaces between
the regions, where electrons have  different
site energies, we find and compare
the transmission coefficients for zigzag, armchair and chiral nanotubes
subjected to the action of an applied step-like  potential.   The correspondence between the nanotube band structure and the energy dependence of the transmission coefficient is demonstrated.
It is shown that the
$(2m,m)$ nanotubes with a medium chiral angle 
reveal   intermediate transport properties as compared with the   achiral armchair, and zigzag nanotubes.

\end{abstract}


\maketitle

\section{Introduction\label{Sec1}}
 
The unique spectroscopic and  electron-transport properties of carbon nanotubes (CNTs)
explain the continuous efforts of experimentalists and theoreticians aimed at the production and investigation of CNT- and graphene-based electronic devices.  Discovered in the early nineties  \cite{Iijima}, the carbon nanotubes are now widely used  in biomedical applications \cite{Bio} and in energy storage facilities \cite{Energy},  thin-film electronics \cite{Films}, as well as in the production of carbon nanotube computers, transistors, and sensors \cite{Liu,sensors, Hills}.

The comprehensive theoretic  
  investigations of the electron and transport properties of CNTs   \cite{Saito,SDD}  started  shortly after their  experimental discovery. 
In recent years, the continued  theoretical investigations of CNTs have been mainly performed  within the Dirac relativistic \cite{Ando,Brey,Peres2} and Schr\"odinger nonrelativistic \cite{Peres1,PRL,PRB,Waka,Saito2017,pss2017,Saroka,Sadykov}   one-particle theories.  In this report, we use the nonrelativistic approach: the analytic solution to the Schr\"odinger equation for the tight-binding H\"uckel-type Hamiltonian.
Our interest is mainly focused
on the electronic structure and transport properties both of the well-studied achiral  armchair and zigzag CNTs,
as well as of the much less investigated chiral nanotubes.   
To be precise, we consider
 medium-chiral-angle $(2m,m)$ CNTs with chirality angle equal to 19.1$^{\circ}$.
 This choice is explained both by the relative simplicity of this structure and by the existence of the well-known and experimentally realized
 procedure of getting  nanotubes with chiral indices
 $(2m,m)$   \cite{Yang,Zhang}.

\begin{figure*}[t!]
 \includegraphics[width=0.9\textwidth]{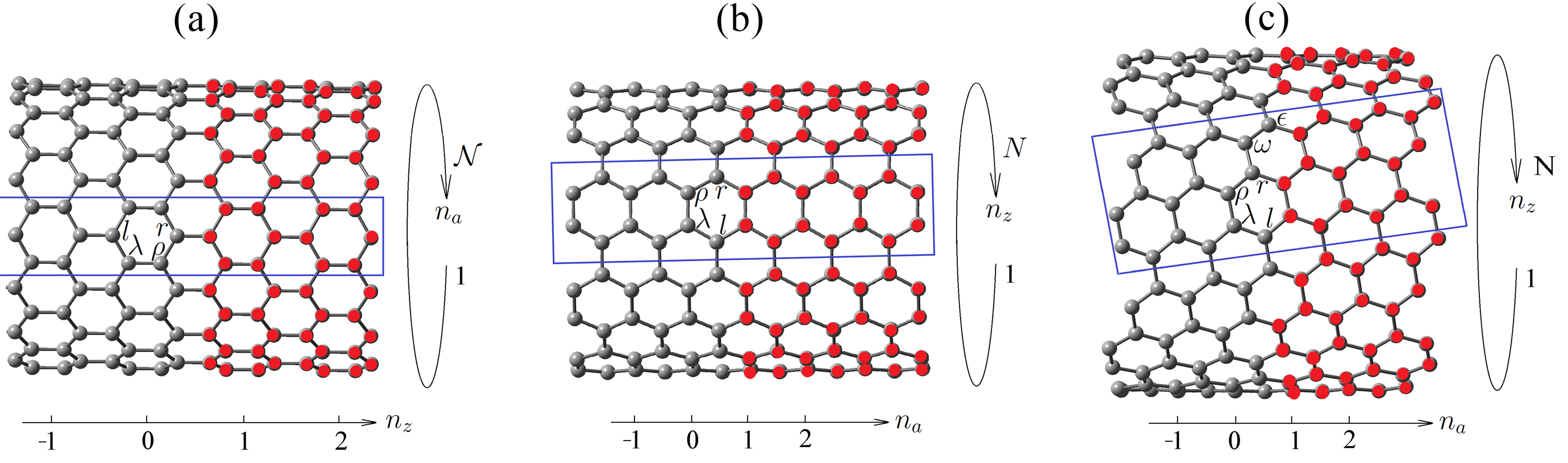}
 \caption{Fragments of infinite zigzag (a), armchair (b) and ($2m,m$) CNTs subjected to the step-like potential.
 The grey and red circles indicate zero and $U$-shifted site energies of carbon atoms.  The labels $ \alpha =\lambda,\rho,  l,r$ for (a), (b) and  $ \alpha =\lambda,\rho,  l,r,\omega,\epsilon$ for (c) are shown for the atoms with $n_z=0$ for (a) and with $n_a=0$ for (b) and (c). The  atoms in the frame have one and the same value of the coordinate $n_a$ in (a) and $n_z$ in (b) and (c).  }
\label{Fig1}
 \end{figure*}

Our goal is to find the transmission coefficients for the achiral and chiral CNTs   in a potential that has
a step-like  profile.
We start with the solution of the
stationary
Schr\"odinger equation ${\mathbf H}\Psi = E\Psi$ with the nearest-neighbor
tight-binding Hamiltonian for  infinite achiral and chiral nanotubes  presented in Fig.~\ref{Fig1}.
We use the one-electron approximation, and the model  represents an ideal $\pi$ electron system. In the nearest-neighbor approximation, 
for the lattice site enumeration explained in Fig.~\ref{Fig1}, the electronic properties of CNTs can be well described by the H\"uckel-type Hamiltonian
\begin{equation}\label{1}
{\mathbf H}= \sum_{\bf r'-r=q}\sum_{\bf r}
[ \varepsilon_{\bf r}(1-\delta_{\bf q,0}) 
-\beta] c^+_{\bf r}c_{\bf r+q},
\end{equation}
where $c^+_{\bf r}$ ($c_{\bf r}$) is the electron creation (annihilation) operator at 
${\bf r} = n_a,n_z,\alpha$; $\varepsilon_{\bf r}$ and $-\beta$ ($\beta>0$) are, respectively, the electron site energy  
and the C--C hopping integral.  
The wave function of $\pi$ electron reads 
$\Psi= \sum_{\bf r}\psi_{\bf r}c^+_{\bf r}|0\rangle$, where
the expansion coefficients obey  the corresponding sets of equations depending on the type of CNTs. 
The Fermi energy of $\pi$ electrons is equal to zero and serves as the reference. In what follows, energy is always expressed in the 
units of $\beta$.  
To solve the problem of  particle transmission under a potential  
with  
 step-like profile, we consider the electron transmission through 
CNTs  with site energies taking the values 0 from the left and $U$ from the right of the interface, as shown in Figs.~\ref{Fig1} a--c.

 It follows
from Fig.~\ref{Fig1} that a unit cell of a CNT contains the atoms with labels $\alpha=l,r,\lambda,\rho$ for the zigzag and armchair CNTs, and
the atoms with $\alpha=l,r,\lambda,\rho, \omega,\epsilon$ for ($2m,m$) CNTs. The  zigzag nanotube (zCT)
can be regarded as a periodic sequence of  polyparaphenylene molecules, the armchair nanotube (aCT)
consist of the acene chains, and the ($2m,m$) CNT is formed by  the acene chains connected with a simple carbon chain (the building blocks of all three CNTs are depicted as blue frames in Fig.~\ref{Fig1}).
As   shown  earlier \cite{PRL,PRB,PSS2021},  the states of $\pi$ electrons in the considered CNTs
can be arranged in $\cal N$ (zCT), $N$ (aCT), and $\mathrm{N}$ (($2m,m$) CNT)  conduction bands and equal number of valence bands, each of which is subdivided into two (zCT and aCT) and three (($2m,m$) CNT) subbands. 
 Due to the orthogonality of  transverse wave functions (see Eqs. (\ref{a2}) and (\ref{a10}) ), there is no interband transition,  and, hence, the problem of finding the transmission coefficient is reduced to the problem of obtaining the scattering amplitudes for each $j$th (zCT) or $\nu$th (aCT and ($2m,m$) CNT), see Fig.~\ref{Fig1}, conduction band. 
Labeling the wave function to the left and right of the interface by $L$ and $R$, respectively, we can write the  
solution to the Shr{\"o}dinger equation  
 for any incident mode $\mu$ as follows (the precise form of these relations   is specified for zCTs, aCTs, and ($2m,m$) CNTs in Sections \ref{Sec2} and \ref{Sec3}):
\begin{align}\label{3}
\psi^{L}_{n_a,n_z,\alpha}& = \psi_{n_a,n_z,\alpha}(\overset{\rightarrow}{k_{\mu}}) + \sum_{\mu_r}
{R}_{\mu_r} \psi_{n_a,n_z,\alpha}(\overset{\leftarrow}{k_{\mu_r}}),\nonumber \\ 
\psi^{R}_{n_a,n_z,\alpha}& =\sum_{\mu_t}{T}_{\mu_t}\psi_{n_a,n_z,\alpha}(\overset{\rightarrow}{\bar k_{\mu_t}}).
\end{align}
 In Eqs. (\ref{3}), the wave vectors are given in $a^{-1}$ units (lattice constant in CNTs $a\approx 2.5\AA$), summation is carried out over  reflected ($\mu_r$) and transmitted ($\mu_t$) modes, and $\psi_{n_a,n_z,\alpha}$ (for any mode) is an  eigenstate. The eigenvalues $E(k)$ are found from the dispersion relation.
The wave  
vector marked by the overbar ($\bar k$), as well as the corresponding eigenenergy $\overline{E}$, refer 
to the nanotubes 
with site energies shifted by $U$: $\overline{E} = E -U$.  In what follows, we consider  only positive energies $E\geq 0$, while the sign of $\overline{E}$ depends on $U$. 

 \section{Achiral (zigzag and armchair) CNTs}\label{Sec2}

The system of equations for the wave functions $\psi_{n_a,n_z,\alpha}$ (the meaning of labels is explained in Fig.~\ref{Fig1}) for both achiral CNTs has the same form
\begin{equation}\label{a1}
\begin{split}
E\psi_{n_a,n_z,l}&= -\psi_{n_a,n_z,\lambda} -\psi_{n_a+1,n_z,\lambda}-\psi_{n_a,n_z-1,r},  \\
E\psi_{n_a,n_z,\lambda} &=-\psi_{  n_a,n_z,l}-\psi_{n_a-1,n_z,l}-\psi_{n_a,n_z,\rho},
  \\
E\psi_{n_a,n_z,\rho}&=-\psi_{n_a,n_z,r}-\psi_{n_a-1,n_z,r} -\psi_{n_a,n_z,\lambda},
  \\
E\psi_{n_a,n_z,r}&=-\psi_{n_a,n_z,\rho} -\psi_{n_a+1,n_z,\rho}-\psi_{n_a,n_z+1,l},  
\end{split}
\end{equation}
while the boundary conditions are different for zigzag and armchair nanotubes. 
We start with the simplest case, a zigzag CNT.

\subsection{Zigzag CNT}
A fragment of a zigzag CNT   is drawn in Fig.~\ref{Fig1}a. 
This nanotube has chiral angle $\theta=0^\circ$. Using the presented notation, the periodic boundary conditions can be written as
 $\psi_{0,n_z,l(r)}=\psi_{{\cal N},n_z,l(r)}$, $\psi_{{\cal N}+1,n_z,\lambda(\rho)}=\psi_{1,n_z,\lambda(\rho)}$, where
$
 n_a=1,\dots,{\cal N},
$
 and
$
n_z=-\infty,\dots,\infty$, which implies the following form of solutions to the system (\ref{a1}) for $ \alpha =\lambda,\rho,  l,r$:
\begin{equation}\label{a2}
\begin{split}
\psi_{n_a,n_z,\alpha} =   {\alpha} e^{i\xi_j n_a}   e^{i\kappa n_z},
 \quad  
  \xi_j  \equiv \dfrac{2\pi j}{\cal N},\;
 j=1,\dots,{\cal N},
\end{split}
\end{equation}
and the wave vector $\kappa$, $0\le k\le\pi$, is in the units of  
$a^{-1}$.
The  electron structure and quantum transport in the zCTs (and their counterparts, armchair nanoribbons) were considered
in a number of works \cite{PRB,PRL,KLA,LPA,KlSh,Yura,PSS2015,Sharma,PSS2018}, and the eigenvalues and eigenstates $\psi_{n_a,n_z,\alpha}$ were found in different representations. The eigenfunctions 
$\psi^{(j)}_{m,n,\alpha}(k)$ can be found by using
different approaches \cite{PRB,Waka,KLA}.
 We use here the following form for the coefficients $\alpha$ in (\ref{a2}):
 \begin{equation}\label{a3}
\begin{split}
l&= \mp \sigma_j \dfrac{z^\pm_j}{E^\pm_j} e^{-i\kappa_j^\pm/2}, \qquad \quad r=1,\\
\lambda &= \pm \sigma_j e^{-i{\pi j}/{{\cal N}}}e^{-i\kappa_j^\pm/2}, \quad \rho=-e^{-i{\pi j}/{{\cal N}}}\dfrac{z^\pm_j}{E^\pm_j},
\end{split}
\end{equation}
and the corresponding dispersion relation (for zero site energy):
\begin{equation}\label{a4}
\begin{split}
(E^\pm_j)^2&=4\cos^2 \dfrac{\pi j}{{\cal N}}+1 \pm 4\sigma_j \cos \dfrac{\pi j}{{\cal N}}   \cos \dfrac{\kappa}{2}=\big | z_j^\pm\big |^2, 
\end{split}
\end{equation}
where
\begin{equation}\label{a5}
z_j^\pm\equiv 2\cos \dfrac{\pi j}{{\cal N}}\pm \sigma_j e^{-i   \frac{\kappa}{2} },
\quad \sigma_j\equiv {\rm sing}({\cal N}/2-j).
\end{equation}
The sign $\pm$ in Eq. (\ref{a4}) marks the "plus" and "minus" branches of the dependence $E_j(\kappa)$.
Note that the mode with the lowest absolute energy belongs
 to the "minus" branch. It is easy to see that for any $j$, there are two positive $\kappa$ (that we denote by $\kappa_j^\pm$) satisfying Eq.(\ref{a4}):
 \begin{equation}\label{a6}
 \begin{split}
 \kappa_j^\pm &= 2\arccos \Big [s\dfrac{(E_j^\pm)^2-4\cos^2\frac{\xi_j}{2}-1} 
 {4 \sigma_j \cos\frac{\xi_j}{2} }\Big ], \\ s&\equiv\left \{ \begin{array}{cl }
 1,& \text{for}\quad E^+_j\\[7pt]
 -1,& \text{for}\quad E^-_j.
 \end{array} \right .
 \end{split}
 \end{equation}
 Evidently, if some $\kappa $ satisfies Eq.(\ref{a4}), the same is true for $-\kappa$. Differentiating $E^\pm_j$ by $\kappa$, we find the group  velocities of the corresponding eigenstates:
 \begin{equation}\label{}
\hbar v_\pm=\frac{dE^\pm_\nu}{d\kappa} 
=\mp \dfrac{1}{E} \left | \cos \dfrac{\xi_j}{2}\right |\sin  \frac{\kappa_j^\pm}{2}.
\end{equation}
  Positive velocity corresponds to  the wave going from  left to  right. Then, for the "plus" energy branch,
  $-\kappa_j^\pm$ is chosen for the incident mode, while $\kappa_j^\pm$
 corresponds to  the reflected wave, and for the "minus" branch the situation is opposite. Thus, for any incident mode $j$, relations (\ref{3}) simplify to 
 \begin{align}\label{a7}
\psi^{L}_{n_a,n_z,\alpha}& = \psi_{n_a,n_z,\alpha}(-s\kappa_{j}^\pm) + 
 {R}_{j}^\pm \psi_{n_a,n_z,\alpha}(s\kappa_j^\pm),\nonumber \\ 
\psi^{R}_{n_a,n_z,\alpha}& = {T}_{j}^\pm\psi_{n_a,n_z,\alpha}(-\bar s \bar \kappa_j^\pm)
\end{align}
(to recall, the wave  
vector marked by the overbar ($\bar \kappa_j^\pm$) refer 
to the zCTs  
with site energies shifted by $U$: $\overline{E}_j^\pm= E_j^\pm-U$.)
Substituting these relations in the first (for $n_z=1$) and last (for $n_z=0$) equations of system (\ref{a1}), we find after some algebra (for details, see \cite{KLA,LPA})
\begin{equation}\label{a8}
\begin{split}
\mathcal{T}_{ j}^\pm&= |{T}_{j}^\pm |^2\dfrac{ E_j^\pm \sin\frac{\bar\kappa_j^\pm}{2} }{\overline{E}_j^\pm \sin\frac{ \kappa_j^\pm}{2} }= 
\dfrac{16\cos^2 \frac{\pi j}{{\cal N}}   \sin \frac{\kappa_j^\pm}{2} \sin \frac{\bar \kappa_j^\pm}{2}   }
{ \Big | -U^2+16 \cos^2 \frac{\pi j}{{\cal N}}Q^\pm\Big |}, \\[7pt]
 Q^\pm&= \left \{\begin{array}{ll}
\sin^2\frac{\kappa_j^\pm+\bar\kappa_j^\pm}{4}, & s=\bar s, \\[7pt]
\cos^2\frac{\kappa_j^\pm+\bar\kappa_j^\pm}{4}, & s=-\bar s,
\end{array}
\right .
\end{split}
\end{equation}
where $\bar s=1$ for $\overline{E}_j^+$ and $\bar s=-1$ for $\overline{E}_j^-$.
The main properties of the transmission coefficient, which is equal to the sum of all transmission amplitudes
of the propagating transverse modes
\begin{equation}\label{a9}
\mathcal{T}(E)=\sum_{j=1}^{\cal N}\big (\mathcal{T}_{j}^++\mathcal{T}_{j}^-\big )
\end{equation}
are discussed in Sec.\ref{Sec3}.

\subsection{Armchair CNT}
A fragment of infinite armchair CNT (aCT), a structure having  chiral angle $\theta=30^\circ$, is drawn in Fig.~\ref{Fig1}b. For this case, the periodic boundary conditions in the transverse direction can be written as
 $\psi_{n_a,0,r}=\psi_{n_a,N,r}$, $\psi_{n_a,N+1,l}=\psi_{n_a,1,l}$, where
$
 n_a=-\infty,\dots,\infty,
$
 and
$
n_z=1,\dots,N$, which implies the following form of solutions to the system (\ref{a1}) for $\alpha =\lambda,\rho,  l,r$:
\begin{equation}\label{a10}
\begin{split}
\psi_{n_a,n_z,\alpha} =   {\alpha} e^{ikn_a}   e^{i\xi_\nu n_z},
 \quad
  \xi_\nu  \equiv \dfrac{2\pi \nu}{N},\;
 \nu=1,\dots,N.
\end{split}
\end{equation}
The energy eigenvalues are found from the condition of solvability of system (\ref{a1}) in the form
\begin{equation}\label{a11}
\begin{split}
(E^\pm_\nu)^2&=4\cos^2 \dfrac{k_\nu^\pm}{2}+1 \pm 4\sigma_\nu \cos \dfrac{k_\nu^\pm}{2}   \cos \dfrac{\xi_\nu}{2}
=\big | z_\nu^\pm\big |^2,
\end{split}
\end{equation}
where
\begin{equation}\label{a12}
z_\nu^\pm\equiv 2\cos \dfrac{k_\nu^\pm}{2}\pm \sigma_\nu e^{-i   \frac{\xi_\nu}{2} },
 \quad \sigma_\nu\equiv {\rm sing}(N/2-\nu).
\end{equation}
The corresponding eigenstates evidently coincide with Eqs. (\ref{a3}) with the interchange of transverse and longitudinal quantum numbers: $\xi_j \leftrightarrow k_\nu^\pm$, $\kappa_j^\pm \leftrightarrow  \xi_\nu$. The same symmetry exists for the eigenvalues, which is clearly seen from the comparison of Eqs. (\ref{a4}) and (\ref{a11}).
However, there is an essential difference between zCT and aCT: for the later, for any energy $E$ and $\nu$, the
values of  $k_\nu$ should be found from the solution to the quadratic equation (\ref{a11}):
\begin{equation}\label{a13}
 2\cos \dfrac{k_\nu^{1,2}}{2}=\left |\sqrt{E^2 -\sin^2\dfrac{\xi_\nu}{2}}\pm\sigma_\nu\cos\dfrac{\xi_\nu}{2}\right |.
 \end{equation}
Relation (\ref{a13}) gives the folllowing interval where $ k_{1,2}$ are real:
\begin{equation}\label{a14}
\sin\dfrac{\xi_\nu}{2} \leq E \leq \sqrt{5+4\sigma \cos \dfrac{\xi_\nu}{2}}.
 \end{equation}
For this energy range, we have or four ($\pm  k_{1,2}$), or two ($\pm  k_{1}$) real values of $k$.
From these $k$, we choose only those for which the velocity is positive (we suppose that the particles moves from left to   right). For  two energy branches
we get two electron velocities:
\begin{equation}\label{a15}
\hbar v_\pm=\frac{dE^\pm_\nu}{dk} 
=-\dfrac{1}{E}\,\sin\dfrac{k^\pm_\nu}{2} 
 \left (2\cos \dfrac{k^\pm_\nu}{2} \pm \sigma_\nu \cos \dfrac{\xi_\nu}{2}\right ).
\end{equation}
For the "$+$" energy branch, the velocity is positive for negative $k$, and vice versa. For the "$-$" energy branch, the velocity is negative for $k$ starting from $-\pi$  up to $k$ satisfying
\begin{equation}\label{a16}
2\cos \dfrac{k}{2} = \sigma_\nu \cos \dfrac{\xi_\nu}{2},
\end{equation}
for which the velocity becomes positive, up to $k=0$. Then for $k>0$, up to $k$ satisfying equality
(\ref{a16}), the velocity is negative, and  for larger $k$ up to $k=\pi$ it is positive.

 Thus, to determine the transmission amplitudes, we first find four solutions to Eq.(\ref{a13}),  choose only real solutions (two or four), and choose  those $k$ for which the velocity   (\ref{a15}) is positive.  They are denoted by $\overset{\rightarrow}{k_1},   \overset{\rightarrow}{k_2} $ and/or   $
 \overset{\rightarrow}{\bar k_1},  \overset{\rightarrow}{\bar k_2}$. If some $k_\nu$ and/or $\bar k_\nu$ are complex, then for the wave going from  left to  right, we choose those $k$ whose imagine parts are positive (then, these evanescent waves vanish  with increasing $n$ and do not participate  in the process of transmission). For reflection, we choose the waves going in the opposite direction:  $\overset{\leftarrow}{k_1}=-\overset{\rightarrow}{k_1},   \overset{\leftarrow}{k_2}=-\overset{\rightarrow}{k_2}$, for which the velocities   (\ref{a15}) are negative (real $k$) or whose imagine parts are negative (complex $k$).

Since for any energy and $\nu$ there are at most two incident modes (numbered by $\nu_0\leq 2$),
solution to 
Eqs. (\ref{a1})  for each $\nu_0$ can be written in a single-mode form 
\begin{equation}\label{a17}
\begin{split}
\psi^{L}_{n_a,n_z,\alpha}& = \psi_{n_a,n_z,\alpha}(\overset{\rightarrow}{k_{\nu}}) + 
\sum_{\mu=1}^2 {R}^\mu_{ \nu}  \psi_{n_a,n_z,\alpha} (\overset{\leftarrow}{k_\mu}),  \\ 
\psi^{R}_{n_a,n_z,\alpha}& = \sum_{\mu=1}^2  {T}^\mu_{ \nu} \psi_{n_a,n_z,\alpha}(\overset{\rightarrow}{\bar k_\mu}).
\end{split}
\end{equation}
Substituting relations (\ref{a17}) in the following system obtained from (\ref{a1}):
\begin{equation}\label{a18}
\begin{split}
E\psi^{L}_{0,n_z,l}&= -\psi^{L}_{0,n_z,\lambda} -\psi^{R}_{ 1,n_z,\lambda}-\psi^{L}_{0,n_z-1,r},  \\
\overline{E}\psi^{R}_{1,n_z,\lambda} &=-\psi^{R}_{ 1,n_z,l}-\psi^{L}_{0,n_z,l}-\psi^{R}_{1,n_z,\rho},
  \\
\overline{E}\psi^{R}_{1,n_z,\rho}&=-\psi^{R}_{1,n_z,r}-\psi^{L}_{0,n_z,r} -\psi^{R}_{1,n_z,\lambda},
  \\
E\psi^{L}_{0,n_z,r}&=-\psi^{L}_{0,n_z,\rho} -\psi^{R}_{ 1,n_z,\rho}-\psi^{L}_{0,n_z+1,l},   
\end{split}
\end{equation}
we get four linear equations for four unknowns ${R}_{\nu}^\mu$, ${T}_{ \nu}^\mu$, $\mu=1, 2$.
They allow us to determine the transmission coefficient
\begin{equation}\label{a19}
\mathcal{T}(E)=\sum_{ \nu=1}^N\big (\mathcal{T}_{ \nu}^1 +\mathcal{T}_{ \nu}^2\big ),
\quad \mathcal{T}_{ \nu}^\mu = \dfrac{v_\nu}{v_{\nu}^\mu}\left|{T}^\mu_{ \nu}\right |^2.
\end{equation}
To elucidate the role of chirality in the electron transport through step-like potential, in the next section we perform
the  derivation of $\mathcal{T}(E)$ for the chiral ($2m,m$) CNTs.

\section{Chiral  ($2m,m$) CNT \label{Sec3}}

The system of equations for the wave functions   for infinite ($2m,m$) CNTs, whose fragment is depicted in Fig.~\ref{Fig1}c, can be written as follows \cite{PSS2021}:
\begin{equation}\label{c1}
\begin{split}
E\psi_{n_a,n_z,l}&= -\psi_{n_a,n_z,\lambda} -\psi_{n_a+1,n_z,\lambda}-\psi_{n_a,n_z-1,\epsilon},  \\
E\psi_{n_a,n_z,\lambda} &=-\psi_{  n_a,n_z,l}-\psi_{n_a-1,n_z,l}-\psi_{n_a,n_z,\rho},
  \\
E\psi_{n_a,n_z,\rho}&=-\psi_{n_a,n_z,r}-\psi_{n_a-1,n_z,r} -\psi_{n_a,n_z,\lambda},
  \\
E\psi_{n_a,n_z,r}&=-\psi_{n_a,n_z,\rho} -\psi_{n_a+1,n_z,\rho}-\psi_{n_a,n_z,\omega},  \\
E \psi_{n_a,n_z,\omega}&=-\psi_{n_a,n_z,\epsilon}-\psi_{n_a-1,n_z,\epsilon} -\psi_{n_a,n_z,r},   \\
E\psi_{n_a,n_z,\epsilon}&=-\psi_{n_a,n_z,\omega}-\psi_{n_a+1,n_z,\omega} -\psi_{n_a,n_z+1,l},
\end{split}
\end{equation}
where
$
 n_a=-\infty,\dots,\infty,
$
 with the periodic boundary conditions
\begin{equation}\label{c2}
 \psi_{n_a,0,\omega}=\psi_{n_a,{\mathrm N},\omega}, \quad \psi_{n_a,{\mathrm N}+1,l}=\psi_{n_a,1,l},
 \, n_z=1,\dots,{\mathrm N}.
\end{equation}
To satisfy these boundary conditions, the solution is sought in the form   (\ref{a10}) with $\alpha =\lambda,\rho,  l,r,\omega,\epsilon$, 
 $\nu=1,\dots,{\mathrm N}$, $\pi\leq k\leq \pi$.
Substituting this solution in (\ref{c1}), we obtain the dispersion relation for ($2m,m$) CNTs  \cite{PSS2021}
in the form of three energy branches:
\begin{equation}\label{c3}
\begin{split}
(E_\nu^\mu)^2&=1+4\cos^2\dfrac{k}{2}+4\cos\dfrac{k}{2}\cos \left (-\dfrac{k}{6}+\dfrac{\xi_\nu}{3} +y_\mu \right )\\
&=\big | z_\nu^\mu \big |^2, \quad  \mu =1, 2, 3, \quad \xi_\nu=\dfrac{2\pi\nu}{{\mathrm N}}, \\
  y_1&=0,\quad y_2=-\dfrac{2\pi}{3},\quad y_3=\dfrac{2\pi}{3},
\\
  z_\nu^\mu  & \equiv  2\cos\frac{k}{2} + e^{-i\eta_\nu^\mu},
\quad \eta_\nu^\mu \equiv-\dfrac{k}{6}+\dfrac{\xi_\nu}{3} +y_\mu.
\end{split}
\end{equation}
It is easy to see that the wave numbers $k$ satisfying Eq.(\ref{c3}) can be found from the following equation 
of the 6th order:
\begin{equation}\label{c4}
\begin{split}
 e^{6ik}&+c_1 e^{5ik} +c_2 e^{4ik}+c_3 e^{3ik}+\overline{c_2} e^{2ik} +\overline{c_1} e^{ik}+1=0,\\
 c_1&=6-3E^2+e^{-i\xi_\nu},\\  c_2&= 3(5-5E^2+E^4)+3e^{-i\xi_\nu}+e^{i\xi_\nu},\\
 c_3&=21-27E^2+9E^4-E^6+6\cos\xi_\nu,
  \end{split}
\end{equation}
where the overbar denotes complex conjugate. 
The corresponding eigenstates can be represented as follows:
\begin{equation}\label{c5}
\begin{split}
 l&=- e^{-i\xi_\nu} e^{i\eta_\nu^\mu} 
 \dfrac{z_\nu^\mu }{ E_\nu^\mu} , \quad \lambda
 =e^{-i\xi_\nu} e^{i\eta_\nu^\mu} e^{-ik/2}, \\
 \rho&=- e^{-i\xi_\nu} e^{2i\eta_\nu^\mu} e^{-ik/2}
 \dfrac{z_\nu^\mu }{ E_\nu^\mu}, \\[5pt]
 r&=e^{-i\eta_\nu^\mu} e^{-ik/2},\quad
 \omega=-  e^{-ik/2}
 \dfrac{z_\nu^\mu }{ E_\nu^\mu},\quad\epsilon=1.
 \end{split}
\end{equation}
The  electron velocity for the $\mu$th energy branch 
is obtained by differentiating Eq.(\ref{c3}):
\begin{equation}\label{c6}
\hbar v_\mu=\frac{dE^\mu_\nu}{dk} 
=\dfrac{1}{E^\mu_\nu} \Big(\dfrac{1}{3}\cos\dfrac{k}{2}\sin \eta_\nu^\mu-\sin k-\sin\dfrac{k}{2}\cos \eta_\nu^\mu  \Big ).
\end{equation}
Thus, the procedure of finding the transmission coefficient is similar to that described in Sec.~\ref{Sec2}: we find six roots of Eq.(\ref{c4}),  choose only real $k$, find to which branch they belong, and,
 to choose the waves going from  left to  right, choose  those $k$ for which the velocity $v_\mu$ (\ref{c6}) is positive. The maximal number of such $k$ is 3. They are denoted by $\overset{\rightarrow}{k_1},   \overset{\rightarrow}{k_2},\overset{\rightarrow}{k_3}$ and/or   $
 \overset{\rightarrow}{\bar k_1},  \overset{\rightarrow}{\bar k_2}, \overset{\rightarrow}{\bar k_3}$. For complex $k$, we choose those  whose imagine parts are positive (then we get the evanescent waves vanishing with increasing $n$ and not participating in transmission). For reflection, we choose the waves going from  right to  left:  $\overset{\leftarrow}{k_1},   \overset{\leftarrow}{k_2},\overset{\leftarrow}{k_3}$, for which $v_\mu$ are negative (real $k$) or whose imagine parts are negative (complex $k$).

As a result, for any $\nu$, there are at most three incident modes, that are numbered by $\mu_0\leq 3$.
For each $\mu_0,$ 
solution to 
Eqs. (\ref{1}) in a single-mode form is:
\begin{align}\label{16}
\psi^{L}_{n_a,n_z,\alpha}& = \psi_{n_a,n_z,\alpha}(\overset{\rightarrow}{k_{\mu_0}}) + 
\sum_{\mu=1}^3 {R}^\mu_{ \nu}  \psi_{n_a,n_z,\alpha} (\overset{\leftarrow}{k_\mu})\\ 
\psi^{R}_{n_a,n_z,\alpha}& = \sum_{\mu=1}^3 {T}^\mu_{ \nu}  \psi_{n_a,n_z,\alpha} (\overset{\rightarrow}{\bar k_\mu}).
\end{align}
Substituting these solutions in the system
\begin{equation}\label{a18}
\begin{split}
E\psi^{L}_{0,n_z,l}&= -\psi^{L}_{0,n_z,\lambda} -\psi^{R}_{ 1,n_z,\lambda}-\psi^{L}_{0,n_z-1,\epsilon},  \\
\overline{E}\psi^{R}_{1,n_z,\lambda} &=-\psi^{R}_{ 1,n_z,l}-\psi^{L}_{0,n_z,l}-\psi^{R}_{1,n_z,\rho},
  \\
\overline{E}\psi^{R}_{1,n_z,\rho}&=-\psi^{R}_{1,n_z,r}-\psi^{L}_{0,n_z,r} -\psi^{R}_{1,n_z,\lambda},
  \\
E\psi^{L}_{0,n_z,r}&=-\psi^{L}_{0,n_z,\rho} -\psi^{R}_{ 1,n_z,\rho}-\psi^{L}_{0,n_z,\omega},   \\
\overline{E}\psi^{R}_{1,n_z,\omega}&=-\psi^{R}_{1,n_z,\epsilon}-\psi^{L}_{0,n_z,\epsilon} -\psi^{R}_{1,n_z,r},
  \\
  E\psi^{L}_{0,n_z,\epsilon}&=-\psi^{L}_{0,n_z,\omega} -\psi^{R}_{ 1,n_z,\omega}-\psi^{L}_{0,n_z+1,l},
\end{split}
\end{equation}
we get six linear equations for six unknowns ${R}_{\nu_0,\nu}^\mu$, ${T}_{\nu_0,\nu}^\mu$, $\mu=1, 2, 3$.
They allow us to determine the transmission coefficient
\begin{equation}\label{a19}
\mathcal{T}(E)=\sum_{ \nu}\big (\mathcal{T}_{ \nu}^1 +\mathcal{T}_{ \nu}^2 +\mathcal{T}_{ \nu}^3\big ),
\quad \mathcal{T}_{\nu}^\mu = \dfrac{v_\nu}{v_{\nu }^\mu}\left|{T}^\mu_{ \nu}\right |^2.
\end{equation}

 \begin{figure*}[t!]
 \includegraphics[width=0.88\textwidth]{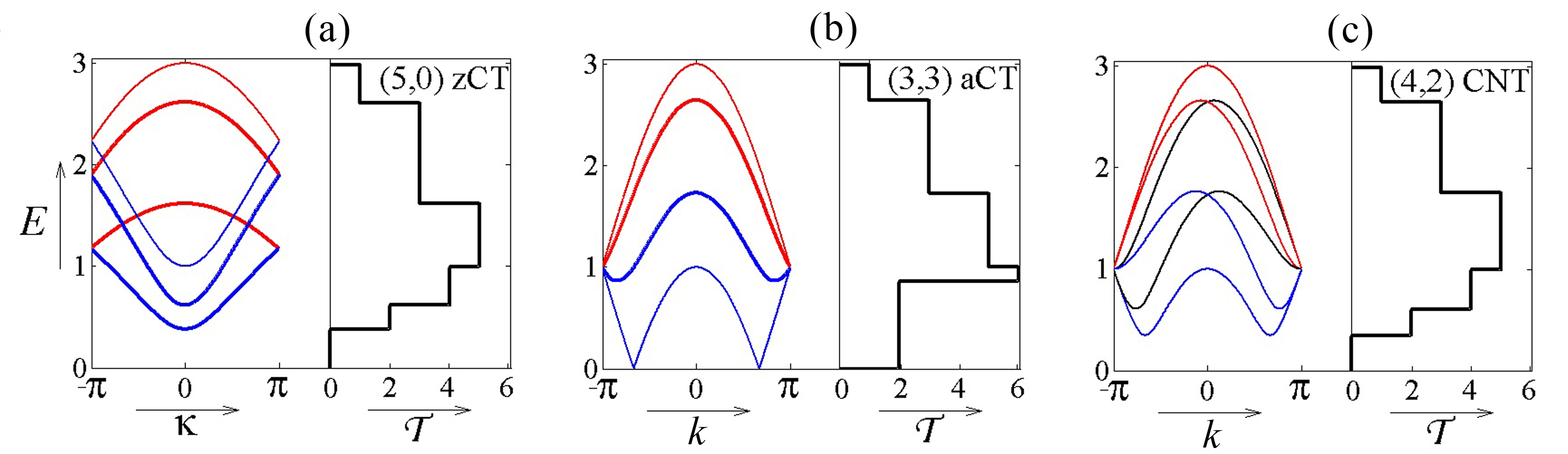}
 \caption{Electron structure and transmission coefficient for, from left to right, zCT (5,0), aCT (3,3) and chiral (4,2) CNT. Double degenerate modes are drawn by thicker lines. Black curves in the right panel: $E_\nu^1$; red curves:  $E_j^+$ (a), $E_\nu^+$ (b), and $E_\nu^2$ (c); blue  curves:  $E_j^-$ (a), $E_\nu^-$ (b), and $E_\nu^3$ (c). }
\label{Fig2}
 \end{figure*}

Now we have at hand all the necessary expressions for the analysis of the properties of transmission coefficients for all three types of CNTs.

 \section{Transmission coefficients}\label{Sec4}

\subsection{Zero potential}
It is instructive to start with the discussion of the dependence $\mathcal{T}(E)$ calculated for $U=0$, when this function is simply equal to the number of "open" modes.
In Fig.~\ref{Fig2}, we plot the dependences $E(k)$ and $E(\kappa)$ and, on the same panel, the corresponding dependences $\mathcal{T}(E)$. 

{\bf zCTs.} For these CNTs, as  is well seen in the left panel, that there are no incident modes (and $\mathcal{T}=0$) if  $E<1-2\cos(2\pi/5)=0.38$, 
in accordance with Eqs. (\ref{a4}),  (\ref{a5}) for $\kappa=0$. For larger energies, two modes open, since the mode $j=2$ is doubly degenerate. When $E>|1-2\cos(\pi/5)|$, there are four open modes, and, thus, $\mathcal{T}=4$.
The maximal value of $\mathcal{T}$ is equal to 5, and it starts to decrease for $E>1+2\cos(2\pi/5)=1.62$.

{\bf aCTs.}  Similar trends are observed for the aCT (middle panel), with an essential difference: since the band gap for any aCT is zero, two modes open immediately when $E$ becomes non-zero, and $\mathcal{T}=2$ up to $E=\sin(\pi/3)=0.87$, the minimum of the double degenerate "minus" energy branch  $E_1^-$ (see Eq.(\ref{a14})). When $E$ becomes larger than $\sin(\pi/3)$, four modes provided by this energy branch become open and $\mathcal{T}=6$. For $E>1$, the number of open modes equals 5 (two from $E_1^-=E_2^-$, two from $E_1^+=E_2^+$, and one from $E_3^+$), up to the maximum of $E_1^-$. With further increase of energy the function  $\mathcal{T}(E)$ decreases in a step-like manner, in accordance with the number of open modes.

{\bf ($2m,m$) CTNs.}  All six modes of (4,2) CNT (right panel) are non-degenerate.
As it follows from Eq.(\ref{c3}), the maximal energy value is attained at $k=0$ for the $\mu=2$ subbranch with $\nu=\mathrm N$: $E^{\mu=2}_{\nu=\mathrm N}(k=0)=3$. 
 The energy gap can be estimated as $2E^{\mu=3}_{\nu=1}(k=2\pi/3)\approx 0.7$. Thus, as is seen in Fig.~\ref{Fig2}c, 
  $\mathcal{T}(E)=0$ for $E\lesssim 0.35$. With further increase of $E$, $\mathcal{T}(E)=2$ up to the energy equal to the minimum of $E^3_2$ (or, which is the same, the minimum of $E^1_2$). The  transmission coefficient is maximal, $\mathcal{T}(E)=5$, for the energies between $E=1$ and   the maximum of $E^3_2$ (or $E^1_2$), when all modes are open except $E^1_3$. For larger energies, the step-like decrease of $\mathcal{T}$ for all three structure is rather similar.
  
  Thus, as the common features of transmission in all three structure, we can mention the step-like increase with increase of energy up to $E\approx 1$, or, since we use the hopping integral energy units, $E\approx 2.7$eV. 
  The maximum of $\mathcal{T}$ equals (or is less by 1 then) the whole number of conducting modes.
  The differences in the behavior of $\mathcal{T}$ for zCTs, aCTs, and chiral CNTs can be expected in the energy region $E\lesssim 1$, while for $E\gtrsim 1$, the three conductance ladders become more and more similar with the energy increase. It can be expected that these regularities remain much the same for small potentials, $U<<1$.  In the next section, we will see what happens with $\mathcal{T}$ for moderate values of $U$,  taking $U=0.25$ and $U=0.5$ as examples.

 \begin{figure*}[t!]
 \includegraphics[width=0.88\textwidth]{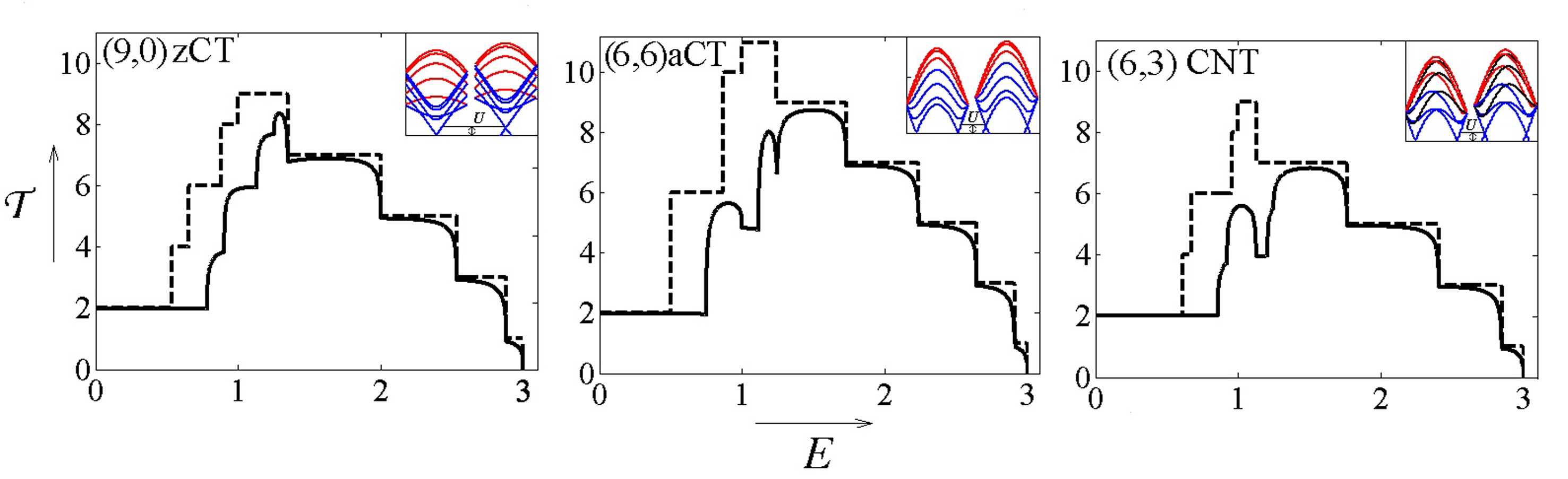}
 \caption{Transmission coefficient  $\mathcal{T}(E)$ for, from left to right, zCT (9,0), aCT (6,6) and chiral (6,3) CNT. Dashed curves: $U=0$, solid curves: $U=0.25$.  Insets: the corresponding dispersion curves (the colors have the same meaning as in Fig/~\ref{Fig2}) for  site energies unshifted (left) and shifted by $U=0.25$ (right). }
\label{Fig3}
 \end{figure*}

\subsection{Non-zero potential}

\begin{figure}[t!]
 \includegraphics[width=0.45\textwidth]{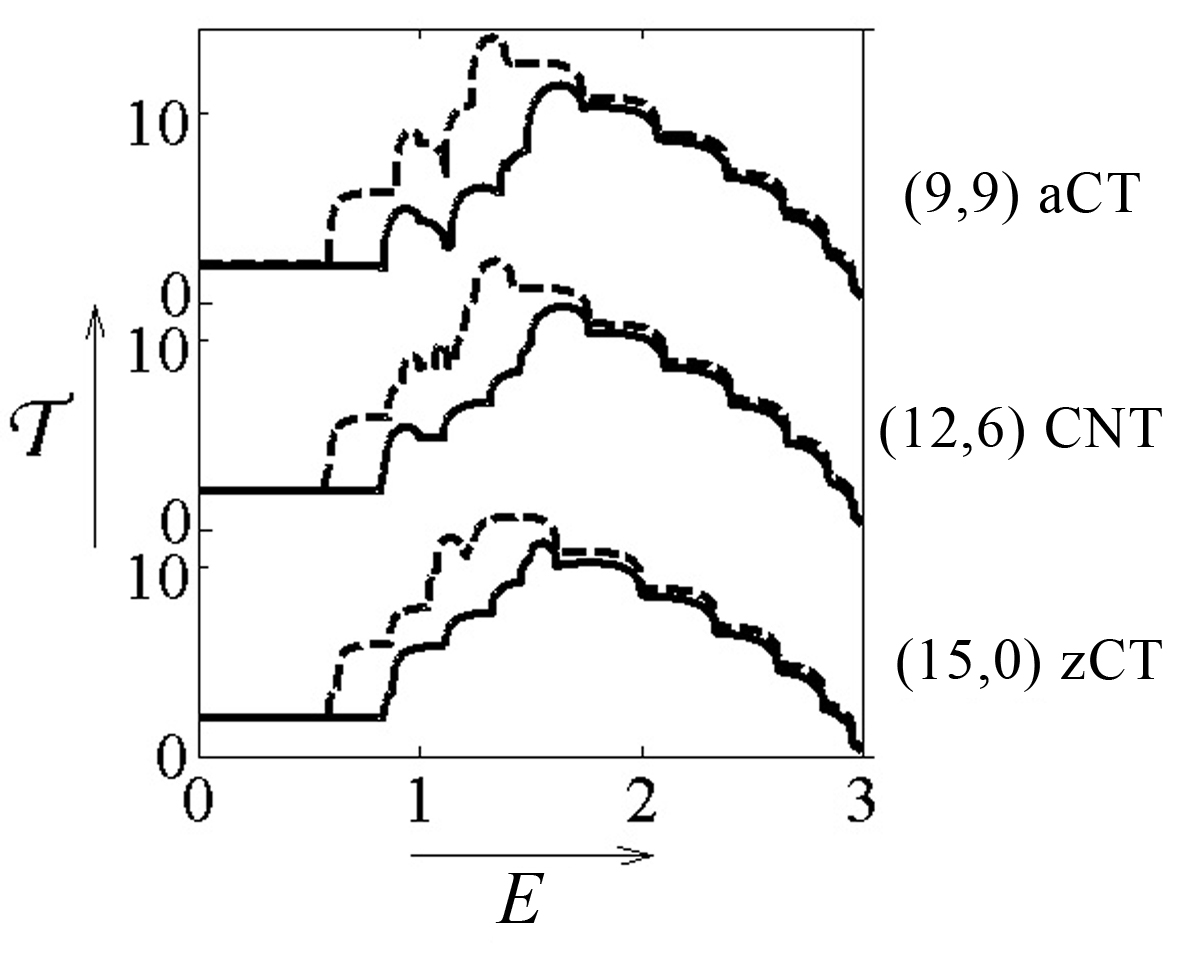}
 \caption{Transmission coefficient  $\mathcal{T}(E)$ for, from bottom  to top, zCT (15,0), chiral (12,6) CNT, and aCT (9,9). All three CNTs have close diameters $d\sim 12\AA$. Dashed curves: $U=0.25$, solid curves: $U=0.5$.    }
\label{Fig4}
 \end{figure}

Due to the symmetry of the problem, we restrict ourselves to the case $U>0$ without loss of generality.
In Fig.~\ref{Fig3} we demonstrate the changes in dependences $\mathcal{T}(E)$ under the step-like potential $U=0.25$ (i.e., $\sim 0.7$ eV). Note that the plots of functions $\mathcal{T}(E)$ for zero potential coincide
with the corresponding dependences calculated in \cite{Chico} with the help of the Green function matching technique,
providing a useful cross-check  on our calculations. Figures~\ref{Fig3} and \ref{Fig4} demonstrate that the transmission coefficient $\mathcal{T}(E)$ is always smaller for larger values of $U$.

As is seen in Fig.~\ref{Fig3}, for all three types of nanotubes considered here, dependences  $\mathcal{T}(E)$ are the same for $E\lesssim 0.5$, namely,
$\mathcal{T}(E)=2$ with a good accuracy for both values $U=0$ and $U=0.25$. This is due to the fact that, as shown in the insets, for moderate values of the potential $U$, exactly two  incident modes belonging to the zero-gap energy band are transmitted to the  same band via two conducting channels. The same behavior of the transmission coefficient for small energies can be expected for all other metallic CNTs.

For large energies, $E \gtrsim 1.5$, we also observe a close similarity of $\mathcal{T}(E)$ for both values of the potential and all presented CNTs.  This can be explained by the simple structure of the electron spectrum
for large energy values: it is seen that the upper subbands of the energy bands have similar bell-like shapes.
At the same time, for energy lying in the interval $0.5 \lesssim E \lesssim 1.5$,
the energy dependences of $\mathcal{T}$ are essentially different for different types of nanotubes. Moreover, as demonstrated in Fig.~\ref{Fig4}  for the CNTs of three considered types and close diameters $d\sim 12\AA$ (to recall, $d=a\sqrt{n^2+m^2+nm}/\pi$ for $(n,m)$ CNTs \cite{SDD}),  the changes in behavior of $\mathcal{T}(E)$ can be organized in the following sequence: zCT $\rightarrow$ ($2m,m$) CNT $\rightarrow$ aCT. This is not at all surprising, taking into account that the corresponding sequence of the chiral angles is $0^\circ \rightarrow 19^\circ \rightarrow 30^\circ$. Thus, the differences in the behavior of electron conductance in the three types of CNT
should be observed    in measurements
of source-to-drain current under varied gate voltage (playing
the role of step potential) only in the energy range $1.3 \div  4$ eV. 
Our calculations show that the increase of the diameter of CNT leads to the smoothness of the dependences of $\mathcal{T}(E)$ and, therefore, reduces the differences
between the process of electron transmission in different nanotubes. Thus, observable variations in $\mathcal{T}(E)$ and, therefore, electron conductance with nanotube types
 can be expected only for rather narrow nanotubes with diameters of up to $30 \AA$.

\section{Conclusions}

In conclusion, we note that we considered the transmission coefficients
for achiral an chiral CNTs of any diameter with the help
 of matching the wave functions on the interface between the regions with zero and nonzero site  
energies.
We used the exact solutions of the model  
Hamiltonian for the description of $\pi$ electron states in ideal carbon achiral and chiral nanotubes.
It is shown that the difference between the dependences of the transmission coefficients for different types of CNTs
is well pronounced within the energy range $1.3 \div  4$ eV for narrow nanotubes with diameters of up to $30 \AA$ 
 and, most likely,  can be observed in the course of the experimental investigations.



\bibliography{Chiral}

\end{document}